# Dual-Channel Reliable Breast Ultrasound Image Classification Based on Explainable Attribution and Uncertainty Quantification

SHUGE LEI§*

UNIVERSITY OF SOUTH CAROLINA

HAONAN HU§

TSINGHUA UNIVERSITY

DASHENG SUN

SHENZHEN HOSPITAL OF PERKING UNIVERSITY

HUABIN ZHANG

BEIJING TSINGHUA CHANGGUNG HOSPITAL, TSINGHUA UNIVERSITY

KEHONG YUAN

TSINGHUA UNIVERSITY

JIAN DAI

TSINGHUA UNIVERSITY

JIJUN TANG

UNIVERSITY OF SOUTH CAROLINA

YAN TONG

UNIVERSITY OF SOUTH CAROLINA

This paper focuses on the classification task of breast ultrasound images and researches on the reliability measurement of classification results. We proposed a dual-channel evaluation framework based on the proposed inference reliability and predictive reliability scores. For the inference reliability evaluation, human-aligned and doctor-agreed inference rationales based on the improved feature attribution algorithm SP-RISA are gracefully applied. Uncertainty quantification is used to evaluate the predictive reliability via the Test Time Enhancement. The effectiveness of this reliability evaluation framework has been verified on our breast ultrasound clinical dataset YBUS, and its robustness is verified on the public dataset BUSI. The expected calibration errors on both datasets are significantly lower than traditional evaluation methods, which proves the effectiveness of our proposed reliability measurement.

## 1 INTRODUCTION

In recent years, medical AI algorithms have achieved remarkably high levels of accuracy and precision [1], particularly in various categories of deep learning algorithms such as medical image classification, medical image segmentation, lesion detection, and medical image feature measurement. Some medical AI algorithms have even surpassed the diagnostic accuracy of human physicians [2], effectively assisting radiologists and other medical imaging professionals in the interpretation of medical images.
However, a major impediment to the widespread adoption of medical AI lies in the lack of reliability assessment in these systems. Systems lacking reliability assessment pose significant safety risks, including bias, uncontrollability, and inconsistency. Even though we may be aware of the overall accuracy of AI models on test data, the inability to determine the reliability of individual results makes it challenging for people to trust a non-transparent system making medical decisions. Such systems are not only untrustworthy but also unaccountable, and they cannot be improved in a targeted manner when issues arise [3].

Breast ultrasound is the primary method for breast cancer screening in China and many other developing countries [4]. If AI-guided breast ultrasound image classification can overcome reliability assessment challenges and be integrated into clinical breast screening applications, it has the potential to significantly reduce the workload of ultrasound specialists and enhance the accessibility of breast cancer screening. Furthermore, as breast ultrasound image classification already has high-performance AI models, with accuracy rates exceeding 90% for benign and malignant classification [5], there are numerous deep learning models to choose from. Research on reliability can expedite the integration of medical AI models into clinical applications.

Studies indicate that trustworthy and reliable AI will gradually become a key driver for industry standardization and commercialization, facilitating the widespread adoption of medical AI. Objective assessments of the reliability of medical AI models have the potential to accelerate the implementation of medical AI while also benefiting research and development, thereby advancing the field of medical AI [6]. Additionally, research in the reliability assessment of medical AI can, to some extent, expand the application of existing explainable methods, fostering further integration and advancement of algorithms from medical AI with those from other domains. Reliable and trustworthy medical artificial intelligence is a common vision in the field of biomedicine. To address the issue of lacking objective reliability assessment methods in the medical AI domain, this paper focuses on the reliability assessment of individual decisions made by medical AI models, with a specific focus on breast ultrasound image classification tasks.

In the context of reliability assessment, this paper is task oriented, centering around breast ultrasound image classification. It employs methods based on feature attribution and uncertainty measurement to evaluate the reliability of deep learning model classifications. This approach enables physicians and patients to make informed decisions regarding the adoption of AI models based on their reliability, thereby providing support for the clinical application of medical image classification models.

## 2 RELATED WORK

### 2.1 Explainable Algorithms

Interpretability is a critical aspect in achieving fair, safe, and reliable AI, emphasizing the need for explanations in a manner understandable by humans. Since the emergence of CAM [7] and LIME [8] in 2016, a series of explainable algorithms have been proposed, known as Explainable AI (XAI). Explainable algorithms can broadly be categorized into attribution methods and linearization methods based on their ultimate effects. Among them, attribution methods are currently the mainstream in explainable algorithm research.

Attribution methods can be fundamentally classified into interpretable networks and post-hoc attribution. Interpretable networks incorporate interpretability into the network's architecture, ensuring the interpretability of the network's inference process. Attribution maps are intermediate products of forward propagation, as seen in models like GAIN [9], CALM [10], SCOUTER [11], and others [12]. Post-hoc attribution, on the other hand, does not require redesigning the network's structure. Instead, it records information such as feature maps and activation maps during the network's inference process and performs feature attribution on the model's output after forward propagation is completed. Since it does not alter the network's structure or parameters, post-hoc attribution has a broader applicability and can be applied to various networks, including those built upon CNN backbones. Examples of post-hoc attribution methods include Grad-CAM [13], Score-CAM [14], Ablation-CAM [15], and others [16]. These methods can be applied to various models using CNN as their backbone.

### 2.2 Uncertainty Measurement

Quantifying uncertainty is challenging in deep networks due to the presence of a large number of parameters. Uncertainty in deep neural networks primarily stems from the uncertainty already present in the data (aleatoric or data uncertainty) and the lack of understanding of the neural network itself (epistemic model uncertainty). The latter can be further divided into homoscedastic (constant across all input data) and heteroscedastic (variable depending on specific inputs). In practice, Bayesian approximation methods such as deep ensembles, Monte Carlo (MC) dropout, and conditional autoencoders are relatively effective algorithms for measuring uncertainty [17].

The most common approach to estimate predictive uncertainty is to measure model uncertainty and data uncertainty separately. Model uncertainty estimate can be simplified by improving the deep neural network, but data uncertainty is inherently complex. Major methods for modeling uncertainty include Bayesian inference methods, ensemble methods, Test Time Augmentation (TTA) methods, and single deterministic networks with explicit components representing model and data uncertainty. Bayesian neural networks apply the full probability theory, allowing the model to compute its uncertainty. However, due to the size of neural network data and parameters, the direct application of approximate Bayesian inference techniques has proven to be challenging. Ensemble methods leverage the synergy between different models to better encapsulate the problem. In addition to improving accuracy, ensemble methods also represent model uncertainty by evaluating the variation in predictions among member models [18]. The fundamental approach to TTA is to create multiple test samples by applying data augmentation techniques to each test sample and then testing all samples to calculate predictive distribution to measure uncertainty. While there are currently various methods available for measuring uncertainty, and some have even applied uncertainty measurement in secondary medical opinions [19], relying solely on uncertainty to assess the reliability of medical AI models is insufficient for a comprehensive reliable evaluation of individual model decisions. Therefore, the design of a comprehensive and holistic framework for assessing the reliability of AI model classifications and establishing baselines remains a significant challenge.

### 2.3 Confidence Calibration

Existing confidence calibration methods aim to provide models with post-hoc confidence correction, aligning their confidence scores more effectively with the true probabilities of correct outcomes. Confidence calibration can broadly be categorized into regularization-based methods [20], Temperature Scaling methods [21], Bayesian methods [23], and methods for generating out-of-distribution (OOD) data [22]. Regularization has a significant impact on network confidence calibration. The mix-up regularization technique involves training by adding interpolated samples of inputs and labels to the dataset, generating new sample-label data by proportionally adding two sample-label data pairs together. Label smoothing is another regularization technique that essentially makes neural networks produce more "flattened" probabilities. Label smoothing can eliminate overfitting, enhancing confidence calibration and reliability. Temperature scaling methods, starting with Platt scaling, use negative log-likelihood loss (NLL) on a validation set for training. Matrix scaling, vector scaling, and temperature scaling are extensions of Platt scaling. MC-Dropout is one of the most widely applied Bayesian methods [23]. MC-Dropout emphasizes applying dropout to model parameters during the inference stage. MC-Dropout is equivalent to randomly sampling from the variational distribution of model parameters, making this integration process more manageable. In image segmentation tasks, histogram binning can be applied to the image's histogram. Bayesian Binning into Quantiles (BBQ) extends binning by performing Bayesian averaging over all possible binning schemes and then using the average number of positive samples in each bin as the calibrated confidence. Introducing adversarial perturbations to input data is beneficial for confidence calibration [24].

## 2.4 Feature Attribution

The field of feature attribution research, which involves attributing features using a model's input and output, has seen limited development. One of the more established methods is RISE, as proposed by Petsiuk et al. [25]. RISE obtains a set of subsets of an input image by randomly masking parts of the input image. The weights for these subsets are determined by the scores corresponding to the predicted class of the original input image. The attribution map is obtained by linearly weighting the subsets. RISE does not impose any requirements on the model structure and only utilizes information from the model, input, and output, making it suitable for various types of models. One of the critical factors influencing feature attribution is the accuracy of feature capture. Objects in images serve as carriers of features, encompassing various semantic features. RISE's random sampling of input images is not based on any semantic information; its sampling units consist of small blocks obtained by uniformly dividing the image. However, the distribution of objects in the image is not uniform, implying that RISE's sampling units may contain incomplete or redundant objects. Developing a post-hoc attribution method that can accurately extract semantic features from images is essential. Super pixels are image subregions composed of adjacent pixels with similar characteristics, such as color, brightness, and texture. These subregions often retain the semantic features of the image and can roughly outline the edges of objects in the image. Super pixels can be seen as a form of dimensionality reduction for image data, transforming pixel-level images into region-level representations. Extracting super-pixels can be accomplished through pixel clustering of the surrounding pixels, as seen in methods like SLIC [26]. Alternatively, they can be obtained through gradient descent-based methods or graph-based methods, such as Turbo pixels [27] and N-cut [28]. In contrast to uniformly divided small grids, super pixels offer a more precise means of extracting objects within images. In the case of breast ultrasound images, the lesion regions within the breast are composed of a collection of super pixels, with each super pixel effectively representing a subset of the lesion area shown as in Figure 1.

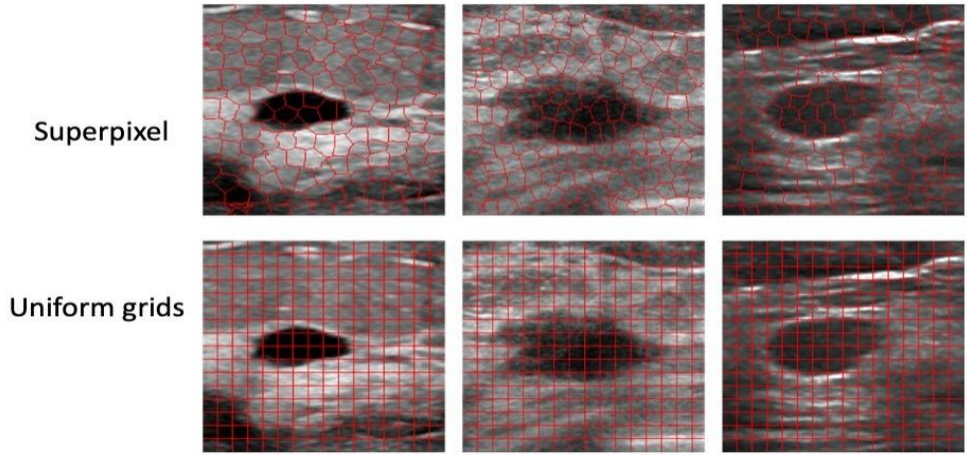

Fig. 1. Effects of super-pixels on breast ultrasound Images

## 3 PROPOSED METHOD

### 3.1 Dual-Channel Reliability Framework

This paper proposes a dual-channel reliability score (DRS) framework as shown in Figure 2, which was computed by the weighted average of the inference reliability score (IRS) and predictive reliability score (PRS) as

$$DRS = \mu \cdot IRS + (1 - \mu) \cdot PRS, \quad (1)$$

while the IRS is computed by the superpixel feature attribution and the PRS is computed by TTA uncertainty measurement.

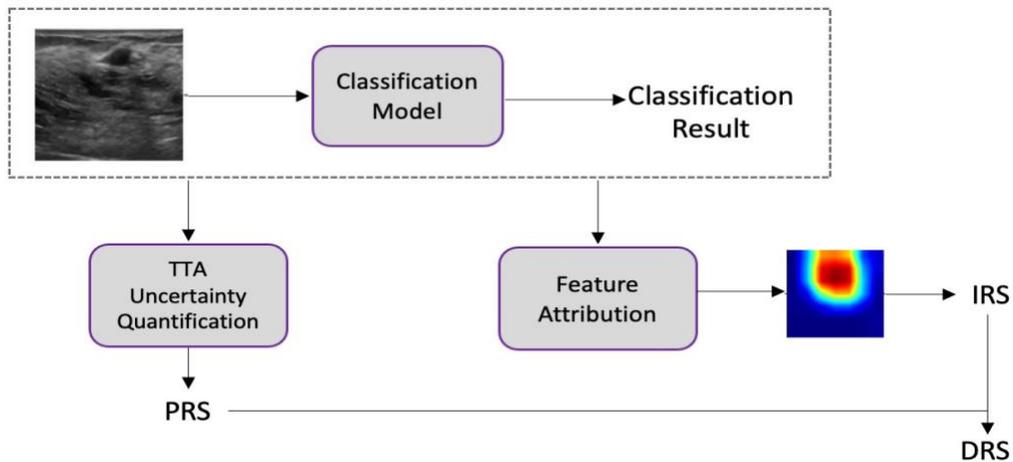

Fig. 2. Dual-channel reliability framework

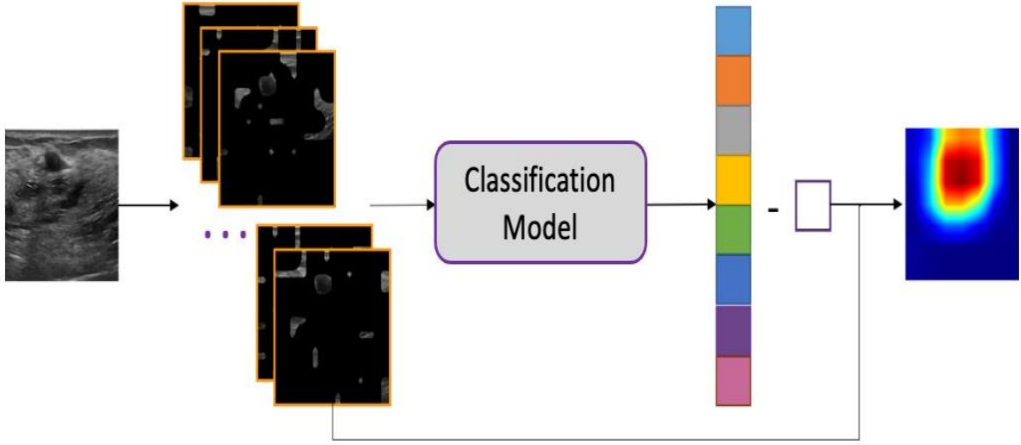

Fig. 3. The framework of SP-RISA

### 3.2 Superpixel Feature Attribution: SP-RISA

A post-hoc attribution method capable of extracting image semantic features was proposed only using model input and output information. This method is named Super-pixel based Random Input Sampling for Attribution (SP-RISA) as shown in Figure 3. When the classification model provides a classification result $y$ for the input image $x$, SP-RISA extracts the superpixels from the input image. Subsequently, it performs $T$ random samplings on the superpixels, resulting in $T$ subsets: $x_1, x_2, \ldots, x_T$, each accompanied by corresponding masks $m_1, m_2, \ldots, m_T$. These subsets $x_1, x_2, \ldots, x_T$ are then fed into the classification model, generating $T$ predictive scores related to class $y$:

$$p_t = G(x_t), \quad t = 1, 2, \ldots, T, \qquad (2)$$

where $p_t$ represents the predicted score for class $y$ corresponding to the $t^{th}$ input subset, and $G$ represents the classification model. By linearly weighting the $T$ masks, the attribution map $A$ is obtained as

$$A = \frac{1}{T} \sum_{t=1}^{T} p_t m_t. \qquad (3)$$

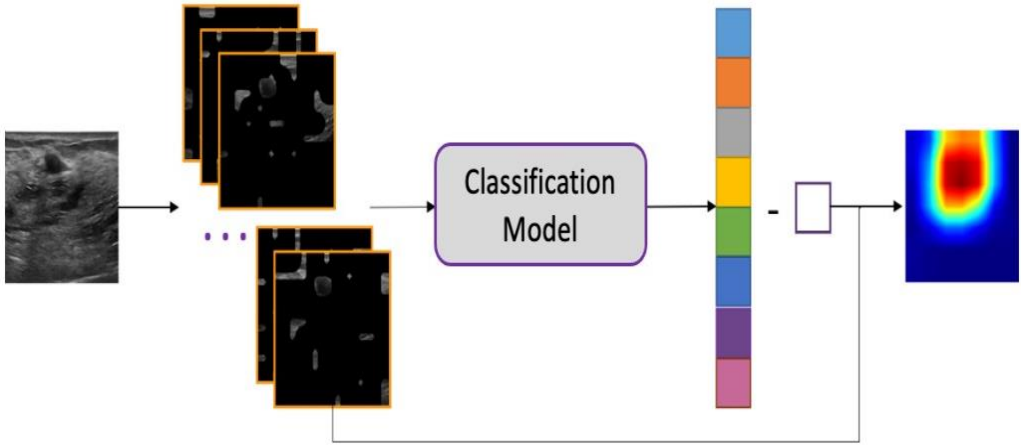

Fig. 3. The framework of SP-RISA

### 3.3 Inference Reliability

From [29], based on the overlap of the model's interest and the instances, the model's inference rationales can be classified into eight categories. These categories include Human Aligned, Sufficient Subset, Sufficient Context, Context Dependent, Confuser, Insufficient Subset, Distractor, and Context Confusion. These eight scenarios can be described using three metrics: Intersection over Union as in Eq.(4), Ground Truth Coverage as in Eq.(5), and Saliency Coverage as in Eq.(6), where $G$ represents the instance region in the image, and $S$ represents the region of interest to the model.

$$IoU = \frac{|G \cap S|}{|G \cup S|} \quad (4)$$

$$GTC = \frac{|G \cap S|}{|G|} \quad (5)$$

$$SC = \frac{|G \cap S|}{|S|} \quad (6)$$

The inference criterion for the reliability framework was proposed based on the aforementioned inference rationales, aligning with one of the three doctor-trusted inference rationales: Human-aligned, Sufficient Subset, or Context-dependent as shown in Figure 4. The reasonable inference should primarily cover the great area around the breast lesion area $GT_{pro}$. Therefore, the inference rationale $S$ must satisfy the following conditions:

$$\begin{cases} |S \cap GT_{pro}| > |S - S \cap GT_{pro}| \\ S \cap GT > 0 \end{cases} \quad (7)$$

$GT$ represents the breast lesion area. $GT_{pro}$ represents the union of the breast lesion area, the area surrounding the lesion, and the area below the lesion. A semantic segmentation model [30] was employed to obtain a mask $M$ for the breast lesion region within the input image $x$. $M_{pro}$ is the union mask of the breast lesion area, the area surrounding the lesion (by an enlarging factor of $k$), and the area below the lesion (by shifting downward by $h$ pixels). It is a binary mask with the same dimensions as the input image, where the mask covers lesion regions with pixel values of 1 and leaves the rest with pixel values of 0. According to medical prior knowledge and clinical experience, we set $k$ to 1.21 [31].

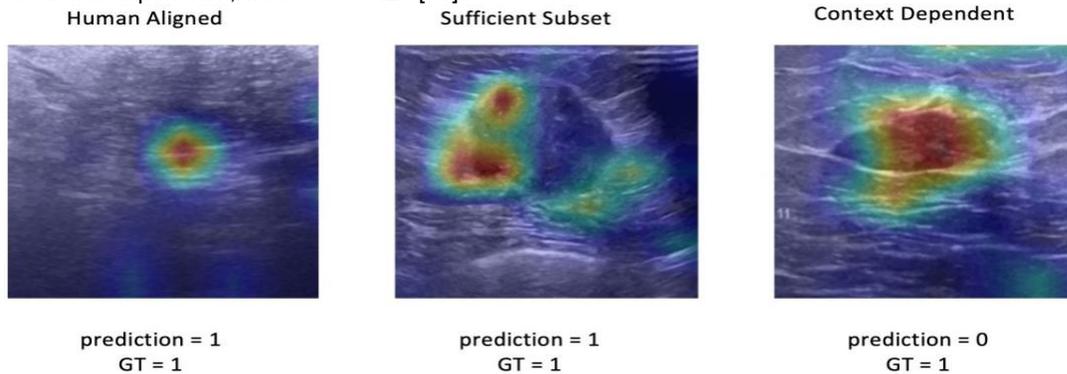

Fig. 4. The doctor-trusted inference Rationales

The inference rationale $S$ is obtained through SP-RISA. The attribution map $A$ was normalized using min-max normalization to obtain $A_n$. An adaptive threshold for binarizing $An$ was employed to ensure that the resulting Eq.(8) is a binary mask of the same size as the mask $M_{pro}$, whereas $threshold(,)$ sets the top $s$ pixels in the image to 1 and the remaining pixels to 0. The inference rationale S intersecting between $G_{pro}$ and $G$, denoted respectively as $I_{pro}$ and $I$, is calculated as Eq.(9) and Eq.(10).

$$S_m = threshold(An, s) \quad (8)$$

$$I_{pro} = \Sigma(S_m \odot M_{pro}) \quad (9)$$

$$I = \Sigma(Sm \odot M) \quad (10)$$

The ratio $E_M$ was used as Eq.(11), which is the energy of $GT$ pixels in the attribution map relative to the total energy of the attribution map.

$$EM = \Sigma(An \odot M). \quad (11)$$

$E_M$ was proved to be no more than 0.5, as shown below:

$$\begin{aligned} E_M &\leq \Sigma(A_n \odot M) \\ &\leq \Sigma(A_n \odot M), \quad when\ I = 0. \\ &= 1 - \Sigma(A_n \odot M) \\ &= 1 - E_M \end{aligned} \quad (12)$$

When the intersection of the inference basis $S$ and $GT$ is not empty, the model's inference process is reliable, and the inference reliability score $IRS$ should be no less than 0.5. In this case, the inference reliability score $IRS$ is calculated as the average of the $IoU$ between the inference basis $S$ and $GT_{pro}$, as follows:

$$IRS = \begin{cases} \frac{I_{pro}}{\Sigma S_m + \Sigma(M_{pro} - I_{pro})}, & I < 0 \\ E_M, & I = 0 \end{cases}. \quad (13)$$

### 3.4 Predictive Reliability

TTA was used for predictive reliability assessment. TTA is a method used for ensemble optimization of model output results [32]. The augmented images, along with the original input image, are fed into the breast ultrasound image classification model for prediction, resulting in a series of classification predictions, denoted as $y_1, y_2, \ldots, y_j$, where $j$ represents the number of adopted data augmentation methods. The proportion $p_i$ of each class was calculated in the $j$ classification results, where $i$ is the class index. The uncertainty score is computed as

$$\mathrm{H} = -\sum_{i=1}^{2} p_i log_{p_i}, \quad (14)$$

where $\mathrm{H}$ is the information entropy. The normalized information entropy $\mathrm{H}$ is used as the Prediction Reliability Score (PRS):

$$PRS = \frac{\mathrm{H}}{\log 2}. \quad (15)$$

## 4 EXPERIMENTS AND RESULTS

### 4.1 Datasets and Metrics

The experiments were conducted on two breast ultrasound datasets. The first one was collected from breast ultrasound examination data at Peking University Shenzhen Hospital, Shenzhen Baoan Maternity and Child Health Hospital, and Beijing Tsinghua ChangGung Hospital, named as the Breast Ultrasound Image Classification Datasets (YBUS). The ultrasound images in the datasets were sourced from 2,805 patients and annotated by five different physicians. The ultrasound devices used for data collection were from four manufacturers: PHILIPS, Mindray, Hitachi, and GE. The datasets contain 3,497 malignant lesion images and 2,061 benign lesion images. The data was divided into training, validation, and test sets in proportions of 80%, 10%, and 10%, respectively. The training and validation sets
were used to train and select breast ultrasound image classification models, while the test set was used to assess the effectiveness of reliability evaluation.

The BUSI dataset comprises breast ultrasound images from 600 female subjects, including three categories: benign, malignant, and normal, totaling 780 images [33]. There are 487 images in the benign category, 210 images in the malignant category, and 133 images in the normal category. Each image is accompanied by corresponding segmentation labels. Both images and labels are in PNG format, with label images in grayscale. Image sizes range from 80KB to 600KB, and the entire dataset is approximately 200MB in size. As this study is focused on breast ultrasound image classification between benign and malignant categories, only benign and malignant data from the BUSI dataset were selected and used.

Considering the low tolerance for false negatives in medical image classification tasks, this study employs recall, accuracy, and F1-score as evaluation metrics for the performance of breast ultrasound image classification models.

For the double-channel reliability assessment framework, this research utilizes Expected Calibration Error (ECE) [20] as the evaluation metric:

$$ECE = \frac{1}{N}\sum_{b=1}^{B} |f_b - acc_b| \cdot S_b, \quad (16)$$

where $N$ represents the total number of samples, $B$ is the total number of bins, $f_b$ represents the average predicted value in the $b_{th}$ bin, $acc_b$ is the accuracy in the $b_{th}$ bin, and $S_b$ is the number of samples in the $b_{th}$ bin. Traditional equal-width binning methods have been shown to introduce significant bias. Furthermore, methods that calculate ECE with a given number of bins exhibit systematic and non-negligible statistical bias. Therefore, this study adopts an automatic search for an equal-sample ECE calculation method that automatically searches for the number of bins that makes the accuracy in each bin monotonically increase and bins the samples according to an equal-sample criterion. Compared to traditional ECE calculation methods, this approach exhibits smaller bias.

## 4.2 Image Feature Attribution Experiment

The SLIC algorithm [26] was employed for superpixel extraction and the extracted superpixels were further used as the fundamental sampling units to implement SP-RISA. The experimental parameters are configured as in Table 1. The effects of SP-RISA are shown in Figure 5. As shown in Figure 6, SP-RISA effectively localizes objects in the image and identifies the key regions of interest to the model.

Table 1: Parameters configuration for feature attribution experiment

| Configure | Parameters |
| --- | --- |
| SLIC Superpixel Area | 30 |
| SLIC Iterations | 10 |
| RISE Sampling Quantity | 5000 |
| SP-RISA Sampling Quantity | 4000 |

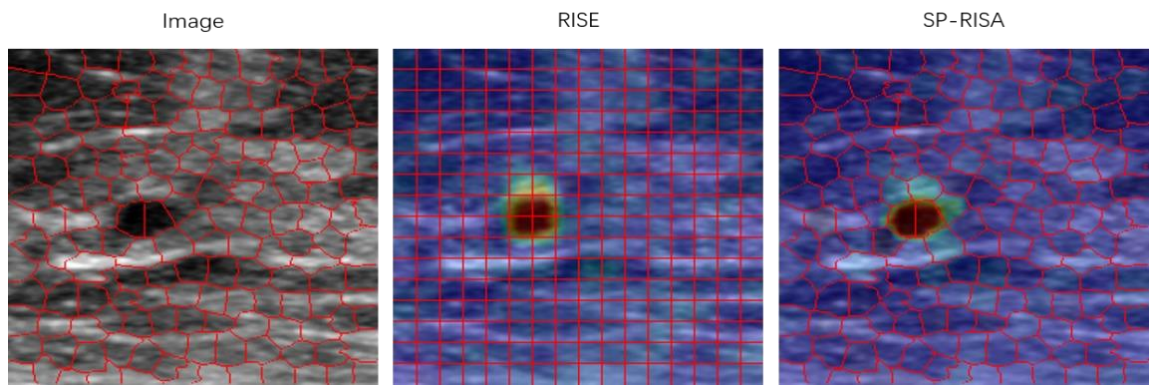

Figure 5: Effects of RASA and SP RISA

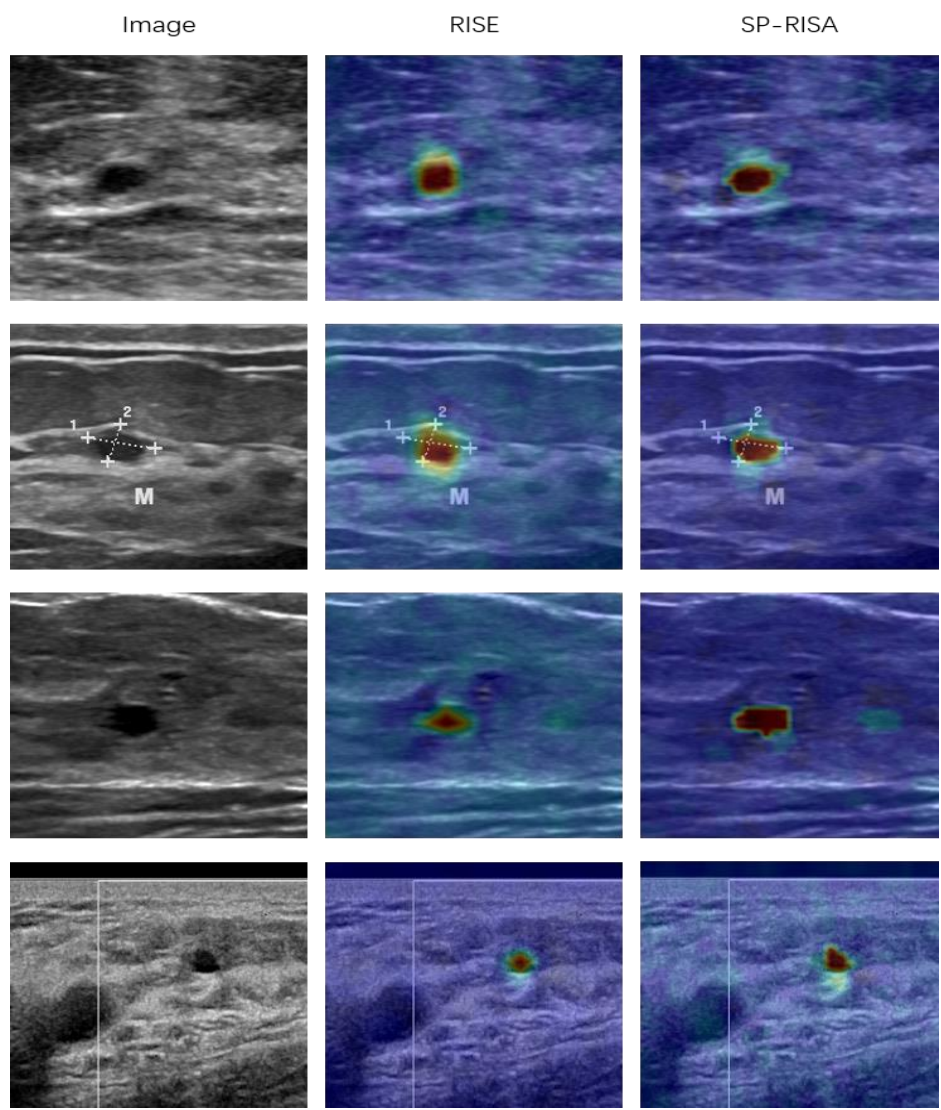

Figure 6: Results of RISE and SP-RISE on breast lesion images

### 4.3 Reliability Experiment Results

The semantic segmentation model employed in the inference reliability assessment module was trained on the BUSI breast ultrasound image dataset [33] using the U-net architecture, as described in [30], achieving a Dice coefficient of 0.99822.
For the dual-channel reliability assessment framework, the accuracy, recall, F1-score, and the average dual-pipe reliability score (mDRS) of VGG16, ResNet50, and ViT_b on the YBUS dataset are presented in Table 2. We use ECE For the reliability assessment. The experimental results are shown in Table 3. A smaller ECE value indicates that the assessed reliability is closer to real-world reliability.

Table 2: Reliability evaluation experiment results

| Model | Precision | Recall | F1 | mDRS |
|---|---|---|---|---|
| VGG16 | 0.83333 | 0.94554 | 0.81425 | 0.88459 |
| ResNet50 | 0.84615 | 0.96164 | 0.83641 | 0.90962 |
| ViT b | 0.89744 | 0.95905 | 0.89045 | 0.85268 |

Table 3: ECE between mDRS and accuracy

| Reliability Measurement | ECE | | |
|---|---|---|---|
|  | VGG16 | ResNet50 | ViT b |
| Confidence | 0.20664 | 0.10561 | 0.14732 |
| 1-Uncertainty | 0.19204 | 0.06771 | 0.09257 |
| mDRS(Ours) | 0.05669 | 0.0586 | 0.07821 |

The results in Table 3 demonstrate that the dual-channel reliability assessment framework designed in this study has ECE values smaller than traditional reliability assessment methods for VGG16, ResNet50, and ViT_b networks. Moreover, it improves ResNet50 by over 40%, with a significant 70% improvement for VGG16, thus confirming the effectiveness of the dual-channel reliability assessment framework.

## 5 CONCLUSION

Current research in the field has seen a limited exploration of reliable medical image classification. This study introduces a framework for reliable measurement and demonstrates promising results on existing datasets and mainstream deep learning models. While this research has made significant strides, there remains limitation and further exploration:

1. The reliability assessment framework comprises inference reliability assessment and prediction reliability assessment. For the prediction reliability assessment, we utilized the TTA method to measure uncertainty. Future research could explore alternative uncertainty measurement methods, such as MC dropout, or delve into the development of data augmentation techniques that balance the preservation of breast ultrasound image fidelity, lesion region integrity, and image diversity. These efforts should aim to create more computationally efficient and accurate prediction reliability assessment methods.

2. The dual-channel reliability score was calculated by taking a weighted average of inference reliability scores and prediction reliability scores. Subsequent research can investigate the incorporation of temperature scaling and other techniques when synthesizing these scores. Additionally, it is essential to explore the impact of different hyperparameters on the effectiveness of reliability assessment and assess whether these hyperparameters maintain low sensitivity when faced with data domain shifts or model variations.

3. The inference reliability for this work is especially costumed for ultrasound imaging readings thus it might not apply to other medical imaging. The framework of the inference reliability can shed light on other applications though.

## ACKNOWLEDGMENTS

The author(s) disclosed receipt of the following financial support for the research, authorship, and/or publication of this article: This research was supported by Foundation of Shenzhen Science and Technology Planning Project (No. GJHZ20200731095205015), the International Cooperation Foundation of Tsinghua Shenzhen International Graduate School (No. HW2021001).

14